\documentclass[amsmath,prb,twocolumn]{revtex4}
\usepackage{color}
\usepackage{amsfonts}
\usepackage{amssymb}
\usepackage{graphicx}
\usepackage{pstricks}

\begin{document}
\title{Quantum Thermodynamics: Inside-Outside Perspective}
\author{Jiayang Zhou}
\affiliation{Department of Chemistry \& Biochemistry, University of California San Diego, La Jolla, CA 92093, USA}
\author{Anqi Li} 
\affiliation{Department of Chemistry \& Biochemistry, University of California San Diego, La Jolla, CA 92093, USA} 
\author{Michael Galperin}
\email{migalperin@ucsd.edu}
\affiliation{Department of Chemistry \& Biochemistry, University of California San Diego, La Jolla, CA 92093, USA} 

\begin{abstract}
We introduce an energy-resolved variant of quantum thermodynamics for open systems strongly coupled to their baths.
The approach generalizes the Landauer-Buttiker inside-outside duality method
[Phys. Rev. Lett. \textbf{120}, 107701 (2018)] to interacting systems subjected to arbitrary external driving.
It is consistent with the underlying dynamical quantum transport description and is capable
of overcoming limitations of the only other consistent approach
[New J. Phys. \textbf{12}, 013013 (2010)].
We illustrate viability of the generalized inside-outside method with numerical simulations for 
generic junction models.   
\end{abstract}

\maketitle

{\bf Introduction.}
Construction of quantum molecular devices became a reality due to the advancement of 
experimental techniques at nanoscale~\cite{jezouin_quantum_2013,pekola_towards_2015,hartman_direct_2018,klatzow_experimental_2019}.
This development poses a challenge to theory making
thermodynamic formulation at nanoscale important from both fundamental and applicational perspectives. 
Indeed, besides purely academic interest, such formulation is at the heart of any efficiency estimate 
of thermoelectric nano-devices~\cite{reddy_thermoelectricity_2007,lee_heat_2013,kim_electrostatic_2014,zotti_heat_2014,cui_perspective:_2017,cui_quantized_2017,cui_peltier_2018,cui_thermal_2019}. 
Existing theories rely on figure of merit concept which stems from 
classical macroscopic formulation. It is restricted to equilibrium considerations (linear response) 
and completely disregards quantum fluctuations. 

Significant progress was achieved in theoretical formulations of quantum thermodynamics 
for systems weakly coupled to their baths. 
Analogs of traditional thermodynamics which focuses on
average system characteristics as well as considerations of quantum fluctuations 
(thermodynamic fluctuation theorems) are available in the literature~\cite{esposito_nonequilibrium_2009,esposito_erratum:_2014}. 

In nanoscale devices, molecule usually forms a covalent bond with contacts on at least one of its interfaces 
which results in hybridization of molecular states with those of the contacts. This appears as system-bath 
interaction of a strength comparable to energy of the isolated system. 
Therefore, thermodynamic formulation of systems strongly coupled to their baths (i.e. situation where
energy of system-bath interaction cannot be disregarded) becomes a practical necessity.
Contrary to weakly coupled situation, thermodynamic formulations for strongly coupled systems
are still at their infancy with most of discussion focused on formulations involving system averages.
Only a few publications on steady-state regime considered quantum fluctuations
in such systems~\cite{esposito_nature_2015}, while fluctuation theorems formulated so far were shown 
to be violated in strongly coupled systems~\cite{brandner_thermodynamic_2018}. 

Here, we focus on thermodynamic theory of averages for systems strongly coupled to their baths
keeping in mind that for future attempts of stochastic thermodynamic formulation for strongly coupled systems
one of the guiding principles of the formulation should be consistency between thermodynamic
and microscopic dynamical descriptions~\cite{kosloff_quantum_2013}.
As far as we know, only one consistent thermodynamic formulation is available in the literature
today. It postulates the von Neumann entropy expression for reduced density matrix 
of the system as proper thermodynamics  entropy. The approach was originally proposed in  Refs.~\onlinecite{lindblad_non-equilibrium_1983,peres_quantum_2002,esposito_entropy_2010} 
and later used in a number of studies~\cite{sagawa_second_2012,kato_quantum_2016,strasberg_quantum_2017,seshadri_entropy_2021}.
The formulation guarantees that in a thermodynamic process which starts from decoupled system and baths
entropy production is positive (i.e. integrated form of the second law of thermodynamics is satisfied) 
while entropy production rate is non-monotonic (i.e. differential form of the second law is not guaranteed)~\cite{esposito_entropy_2010}.
When a thermodynamic process does not start from the decoupled state, 
the second law is not guaranteed in any form.

We argued~\cite{seshadri_entropy_2021} that the deficiency of this von Neumann formulation 
is due to its neglect of energy resolution in system entropy: 
the von Neumann expression operates with reduced density matrix, 
which is a time-local (integrated in energy) object.
Here, we introduce a general energy-resolved thermodynamic formulation for
systems strongly coupled to their baths which is consistent with underlying microscopic dynamics.
This is done by employing nonequilibrium Green's function method in reformulating 
the inside-outside approach in Ref.~\onlinecite{bruch_landauer-buttiker_2018}
originally developed for noninteracting systems under adiabatic driving.
We extend the theory to interacting systems under arbitrary driving.
The resulting formulation
is capable of overcoming limitations of the von Neumann formulation in Ref.~\onlinecite{esposito_entropy_2010}.

\begin{figure}[htbp]
\centering\includegraphics[width=0.8\linewidth]{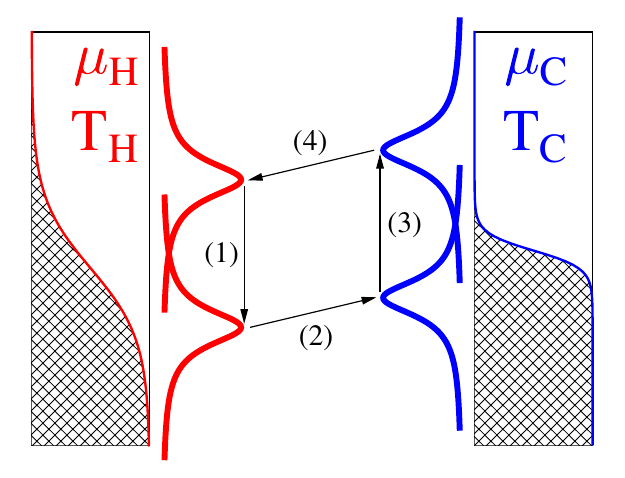}
\caption{\label{fig1}
(Color online)
Sketch of the Carnot cycle in the resonant level junction model.
}
\end{figure}

\newpage
{\bf Model.}
First, we introduce a generic model of open quantum system and mention several concepts from
its microscopic dynamics (quantum transport) description which will be necessary 
for thermodynamic formulation.

We consider a system $S$ strongly coupled to a number of baths 
$\{B\}$ and subjected to arbitrary external driving applied to the system and system-baths 
couplings. Hamiltonian of the model is
\begin{equation}
\label{Ht}
\hat H(t) = \hat H_S(t) + \sum_B\left(\hat H_B + \hat V_{SB}(t)\right)
\end{equation}
where Hamiltonian of the system $\hat H_S(t)$ contains any intra-system interactions, and
\begin{equation}
\begin{split}
\hat H_B &= \sum_{k,\alpha\in B} \varepsilon_{k\alpha}\hat c_{k\alpha}^\dagger \hat c_{k\alpha}
\\
\hat V_{SB}(t) &= \sum_{m\in S}\sum_{k,\alpha\in B}\left(V_{m,k\alpha}(t) \hat d_m^\dagger\hat c_k
+\mbox{H.c.}\right)
\end{split}
\end{equation}
describe bath $B$ and its coupling to the system.
Here, $\hat d_m^\dagger$ ($\hat d_m$) and $\hat c_{k\alpha}^\dagger$ ($\hat c_{k\alpha}$)
creates (annihilates) an electron in orbital $m$ of the system $S$ and state $k$ in channel $\alpha$
of the bath $B$, respectively. 

For description of the system microscopic dynamics we employ
the nonequilibrium Green's function (NEGF) method~\cite{haug_quantum_2008,stefanucci_nonequilibrium_2013},
which is routinely employed in quantum transport formulations for open nanoscale systems.
Thermodynamic formulation below requires several basic concepts of the quantum transport theory.
In particular, we utilize the
single-particle Green's function defined on the Keldysh contour,
\begin{equation}
\label{GF}
G_{m_1m_2}(\tau_1,\tau_2) \equiv -i\langle T_c\, \hat d_{m_1}(\tau_1)\, \hat d_{m_2}^\dagger(\tau_2)\rangle
\end{equation}
(here and below $e=k_B=\hbar=1$),
and expressions for particle $I_B^{N}(t)$ and energy $I_B^{E}(t)$ fluxes at the interface with bath $B$~\cite{jauho_time-dependent_1994,galperin_heat_2007}
\begin{equation}
\label{I_NE}
\begin{split}
I_B^N(t) &=\sum_{\alpha\in B}\int\frac{dE}{2\pi}\, i_\alpha(t,E)
\\
I_B^E(t) &=\sum_{\alpha\in B}\int\frac{dE}{2\pi}\, E\, i_\alpha(t,E)
\end{split}
\end{equation}
Here,
\begin{align}
\label{ia_tE}
i_\alpha(t,E) &\equiv -2\, \mbox{Im}\sum_{m,m_1\in S}\int_{-\infty}^t dt_1\, e^{-iE(t_1-t)}
\\ & \times
2\pi\rho_\alpha(E)\, V_{m_1,\alpha}(E,t_1)\, V_{\alpha,m}(E,t)
\nonumber \\ &\times
\bigg(G_{mm_1}^{<}(t,t_1)+f_\alpha(E)\, G_{mm_1}^{r}(t,t_1)\bigg)
\nonumber
\end{align}
is the unitless energy-resolved particle flux in channel $\alpha$,
$T_c$ is the contour ordering operator, $f_\alpha(E)$ is the
Fermi-Dirac distribution in channel $\alpha$, 
$<$ and $r$ are the lesser and retarded projections of the system Green's function (\ref{GF}), 
$\rho_\alpha(E)$ is the density of states in channel $\alpha$, and $V_{m,k\alpha}(t)=V_{m,\alpha}(\varepsilon_{k\alpha},t)$. 
Finally, we use expression for heat flux $\dot{Q}_B(t)$
from the quantum transport theory~\cite{esposito_nature_2015}
\begin{equation}
\label{QB}
\dot{Q}_B(t)=\sum_{\alpha\in B}\int\frac{dE}{2\pi}\left(E-\mu_B\right)\, i_\alpha(t,E)
\end{equation}
With these definitions we are ready to discuss thermodynamic formulation for systems strongly coupled
to their baths.


{\bf Quantum thermodynamics.}
For systems strongly coupled to their baths the main problem is formulation of the second law of thermodynamics
in terms of quantities used in dynamical description (fluxes and populations of states).
Differential form of the second law is
\begin{equation}
\label{second_law}
\frac{d}{dt} S(t) = \sum_B \beta_B\dot{Q}_B(t) + \dot{S}_i(t)\qquad \dot{S}_i(t)\geq 0
\end{equation}
where $S(t)$ is system entropy, $\dot{Q}_B(t)$ is heat flux at system interface with bath $B$,
and $\dot{S}_i(t)$ is entropy production. In this expression only heat flux is clearly defined by
the microscopic theory, Eq.(\ref{QB}).

To define entropy and entropy production we employ the following observations
from the quantum transport theory:
\begin{enumerate}
\item Overall (system plus baths) dynamics is unitary. Thus, entropy of the universe
given by the von Neumann expression for the total density operator $\hat\rho(t)$,
$S_{tot}(t)\equiv-\mbox{Tr}\left[\hat\rho(t)\,\ln\hat\rho(t)\right]$, does not change
during evolution: $\frac{d}{dt} S_{tot}(t)=0$.
\item Particle fluxes from baths into system are thermal. Non-thermal fluxes from system into baths
do not return into the system. Indeed, within the NEGF dynamics of the system is governed
by the Dyson equation - equation-of-motion for the Green's function (\ref{GF}).
This dynamical law is exact.
Effect of the baths enters the Dyson equation through corresponding self-energies
which only contain information on thermal distribution in the baths.
\item Thermalization processes take place far away from the system. They do not affect
the system dynamics.
\end{enumerate}
Below we use these observation to develop thermodynamic formulation for systems strongly coupled to their baths.

The first observation allows to define system entropy $S(t)$ from total baths entropy $S_{B,tot}(t)$ as~\footnote{We note that Eq.(\ref{Sdef}) is not a manifestation of additivity of system and baths contributions
to the total entropy (due to entanglement the contributions are clearly not additive);
rather it is definition of the system entropy.
That is, one considers evolution of the total universe (system plus baths) and separately evolution
of the baths alone; difference between results of the two evolutions is defined as a quantity characterizing
system entropy. Please note that such way of defining properties of systems strongly coupled to baths
goes back to works by Kirkwood~\cite{kirkwood_statistical_2004} and was used in many previous thermodynamic
considerations~\cite{talkner_colloquium_2020,bruch_quantum_2016,ochoa_energy_2016,bruch_landauer-buttiker_2018}.}
\begin{equation}
\label{Sdef}
 \frac{d}{dt} S_{tot}(t) \equiv \frac{d}{dt}S(t) + \frac{d}{dt} S_{B,tot}(t) = 0
\end{equation}
Defining system quantity from baths characteristics is at the 
heart of {\em the inside-outside approach} first introduced in Ref.~\onlinecite{bruch_landauer-buttiker_2018}.
Original (Landauer-Buttiker) formulation of Ref.~\onlinecite{bruch_landauer-buttiker_2018} is restricted to
noninteracting systems under adiabatic driving. Here, we generalize it
to interacting systems and to arbitrary driving, thus providing 
a general thermodynamic formulation applicable in any transport regime and for any system.
Technically, the generalization requires employing the NEGF method in place of single-particle 
scattering theory, utilizing independent current-carrying states in the baths in place of
scattering states, and accounting for correlations between states of different energies
in addition to inter-channel correlations. 

The second observation allows to separate total particle flux into incoming thermal, $\phi^{in}(E)$,
and outgoing non-thermal, $\phi^{out}(t,E)$, contributions
\begin{equation}
i_\alpha(t,E) = \phi^{in}_\alpha(E) - \phi^{out}_\alpha(t,E)
\end{equation}
Here, $i_\alpha(t,E)$ is energy-resolved particle flux defined in Eq.(\ref{ia_tE}),
$\phi^{in}_{\alpha}(E)$ is thermal population of incoming state in channel $\alpha$ of the baths,
and $\phi^{out}_\alpha(t,E)$ is non-thermal population of outgoing state in channel $\alpha$ of the baths.
To account for system-induced coherences between states of different energies and 
different channels in the baths, in addition to populations one has to consider also coherences
in the energy and channel spaces: $\phi^{out}_{\alpha\beta}(t;E_\alpha,E_\beta)$.
Their explicit form can be obtained from equation-of-motion for the baths density matrix 
(see Appendix~\ref{appA} for derivation)
\begin{equation}
\label{def_phi}
\begin{split}
& \phi_{\alpha\beta}^{in}(E_\alpha,E_\beta) \equiv 2\pi\delta_{\alpha,\beta}\,\delta\left(E_\alpha-E_\beta\right) f_\alpha(E_\alpha)
\\
& \phi_{\alpha\beta}^{out}(t;E_\alpha,E_\beta) \equiv 
\phi_{\alpha\beta}^{in}(E_\alpha,E_\beta)
\\ &
-i(2\pi)^2\rho_\alpha(E_\alpha)\rho_\beta(E_\beta)\sum_{m,m_1\in S}\int_{-\infty}^{t} dt_1\,
\\ &
\bigg(
e^{-iE_\beta(t_1-t)}\, V_{\alpha,m}(E_\alpha,t)\, V_{m_1,\beta}(E_\beta,t_1)
\\ &\quad\times
\left[ G^{<}_{mm_1}(t,t_1)+G^{r}_{mm_1}(t,t_1)\, f_\beta(E_\beta)\right]
\\ & \,\,\, +
e^{+iE_\alpha(t-t_1)}\, V_{\alpha,m_1}(E_\alpha,t_1)\, V_{m,\beta}(E_\beta,t)
\\ &\quad\times
\left[ G^{<}_{m_1m}(t_1,t)-G^{a}_{m_1m}(t_1,t)\, f_\alpha(E_\alpha)\right]
\bigg)
\\ & - (2\pi)^2 \rho_\alpha(E_\alpha)\rho_\beta(E_\beta) (E_\alpha-E_\beta) 
G^{<}_{E_\alpha,E_\beta}(t,t)
\end{split}
\end{equation}
Here, $\rho_{\alpha}(E_\alpha)$ and $\rho_{\beta}(E_\beta)$ are densities of states of
channels $\alpha$ and $\beta$ of the baths, respectively. 
While our approach is general, at steady-state Eq.(\ref{def_phi}) 
reduces to the Landauer-Buttiker scattering theory result.

Having expressions for populations and coherences, Eq.(\ref{def_phi}), one can define
rate of entropy change in the baths as difference between entropy flux
outgoing from the system and entropy flux incoming to the system
\begin{equation}
\label{SB_tot}
\frac{d}{dt}S_{B, tot}(t) \equiv \mbox{Tr}_{c,E}\bigg\{\sigma\left[\mathbf{\phi}^{out}(t)\right]-\sigma\left[\mathbf{\phi}^{in}\right]\bigg\} 
\end{equation}
Here, $S_{B, tot}$ is the total entropy of all the baths, 
$\mbox{Tr}_{c,E}\{\ldots\}$ is trace over channels and energies in the baths, 
and 
\begin{equation}
\mathbf{\sigma}\left[\mathbf{\phi}\right] \equiv -\mathbf{\phi}\,\ln(\mathbf{\phi}) 
- (\mathbf{I}-\mathbf{\phi})\,\ln(\mathbf{I}-\mathbf{\phi})
\end{equation}
is the von Neumann expression for entropy in the baths.

The third observation indicates that neither thermalization process affects physics of the system, 
nor system participates in the process. Thus, entropy production, which takes place during thermalization, 
can be modeled as a process of reducing non-thermal distribution over bath states $\phi^{out}$
to thermal distribution $\phi^{in}$.
Following Ref.~\onlinecite{vedral_role_2002} entropy production can be rationalized as 
information erasure due to measurement of non-thermal baths by set of thermal super-baths
weakly coupled to the baths. 
The process leads to entropy production in the universe which consists of entropy change in the baths,
$\mbox{Tr}_{c,E}\left\{\sigma\left[\phi^{in}\right]-\sigma\left[\phi^{out}(t)\right]\right\}$, 
and heat flux into the super-baths, $-\sum_B\beta_B\dot{Q}_B(t)$. 
Adding the two contributions leads to expression for entropy production
\begin{equation}
\label{dotSi}
\begin{split}
& \dot{S}_i(t) \equiv 
\mbox{Tr}_{c,E}\bigg\{\phi^{out}(t)\,\big[\ln\phi^{out}(t)-\ln\phi^{in}\big]
\\ &
+\left(\mathbf{I}-\phi^{out}(t)\right)\,\big[\ln\left(\mathbf{I}-\phi^{out}(t)\right)-\ln\left(\mathbf{I}-\phi^{in}\right)\big]
\bigg\}
\end{split}
\end{equation}
Note that contrary to formulation of Ref.~\onlinecite{esposito_entropy_2010}
entropy production rate in the inside-outside approach is always positive~\cite{sagawa_second_2012,esposito_quantum_2015}.

Finally, using (\ref{SB_tot}) and (\ref{dotSi}) in (\ref{Sdef}) leads to the
differential form of the second law of thermodynamics for entropy of the system $S(t)$, Eq.(\ref{second_law})
(see Appendix~\ref{appB} for derivation).

This completes thermodynamic formulation for system strongly coupled to its baths.
Note that the introduced formulation readily yields access to energy-resolved version of the 
second law as discussed in Ref~\onlinecite{seshadri_entropy_2021}.

\begin{figure}[t]
\centering\includegraphics[width=\linewidth]{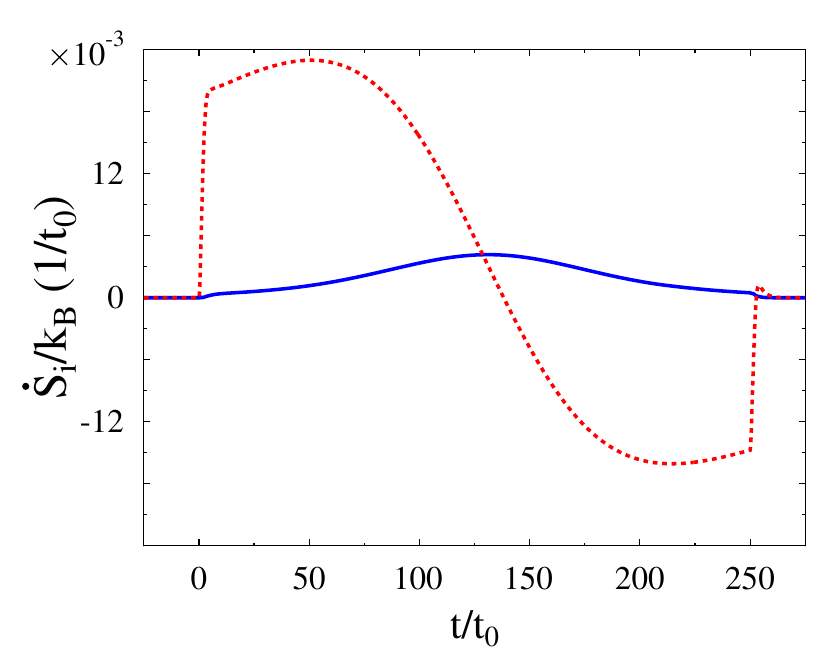}
\caption{\label{fig2}
(Color online) Entropy production rate for an isothermal process in resonant level model
with the level shifted from $0.2$ to $-0.2$. Calculations are done
within the inside-outside approach (see Eq.(\ref{dotSi}), solid blue line)
and within the von Neumann approach of Refs.~\onlinecite{lindblad_non-equilibrium_1983,peres_quantum_2002,esposito_entropy_2010}
(see Eq.(9) of  Ref.~\onlinecite{seshadri_entropy_2021}, dotted red line).
See text for parameters.
}
\end{figure}

{\bf Numerical results.}
We now illustrate the inside-outside approach for the resonant level model
of phonon assisted tunneling where a single molecular level $\varepsilon_0$ represents 
a junction connecting two electron reservoirs ($L$ and $R$) while being also coupled 
to a single harmonic mode $\omega_0$.
The mode represents molecular vibration and is coupled to a thermal (phonon) bath ($P$).
Hamiltonian of the system is given in Eq.(\ref{Ht}) with
\begin{equation}
\hat H_S(t) = \varepsilon_0(t)\hat d^\dagger\hat d + \omega_0\hat a^\dagger\hat a
+ M\left(\hat a+\hat a^\dagger\right)\hat d^\dagger\hat d
\end{equation}
and $B\in\{L,R,P\}$ where
\begin{equation}
\begin{split}
\hat H_K &=\sum_{k\in K}\varepsilon_k\hat c_k^\dagger\hat c_k\qquad (K=L,R)
\\
\hat V_{SK}(t)&=\sum_{k\in K}\left(V_k(t)\hat d^\dagger\hat c_k+V_k^{*}(t)\hat c_k^\dagger\hat d\right)
\\
\hat H_P &= \sum_{p\in P}\omega_\alpha\hat b_p^\dagger\hat b_p
\\
\hat V_{SP} &= \sum_{p\in P}\left(U_p\hat a^\dagger\hat b_p+U_p^{*}\hat b_p^\dagger\hat a\right)
\end{split}
\end{equation}
Driving is performed in the position of the level, $\varepsilon_0(t)$, 
and in system-fermionic baths coupling strengths, $V_{k}(t)\equiv u(t) V_k$.
To simplify simulations we assume the wide-band approximation (WBA)
which allows to reduce the general expressions (\ref{def_phi}) to
energy diagonal form (see Appendix~\ref{appC}).
Also, incorporation of the bosonic bath $P$ requires slight generalization of
the thermodynamic formulation (see Appendix~\ref{appD}).
Electron-phonon interaction $M$ is treated within the self-consistent Born approximation
(SCBA)~\cite{park_self-consistent_2011}.

Parameters of the simulation are given in terms of arbitrary unit of energy $E_0$
and corresponding unit of time $t_0=2\pi/E_0$. 
Unless stated otherwise, the parameters are as follows.
Vibrational frequency $\omega_0=0.05$,
electron-phonon interaction $M=0.01$,
temperature $T=0.05$, electron escape rate 
$\Gamma_0\equiv 2\pi\sum_{k\in K}\lvert V_k\rvert^2\delta(E-\varepsilon_k)=0.1$, 
and energy dissipation rate $\gamma(\omega)\equiv 2\pi\sum_{p\in P}\lvert U_\alpha\rvert^2\delta(\omega-\omega_\alpha)=\theta(\omega)\,\gamma_0\,\frac{\omega^2}{\omega_0^2} exp(1-\omega/\omega_0)$, where $\gamma_0=0.1$. 
The junction is not biased, the Fermi energy is taken as origin, $E_F=0$, 
and the temperatures in the fermionic and bosonic baths are assumed to be the same.
Simulations are performed on 
energy grid with $4001$ points spanning a range from $-2$ to $2$ with step size $10^{-3}$.
FFTW fast Fourier transform library~\cite{FFTW} was employed in the simulations
(see Appendix~\ref{appE} for details).

Figure~\ref{fig2} shows entropy production rate for the resonant level  coupled to a reservoir
and driven from $0.2$ to $-0.2$ position with the constant rate 
$\dot{\varepsilon_0}=1.6\times 10^{-3}\, E_0/t_0$.
The simulation is performed in the absence of electron-phonon coupling, $M=0$.
We compare our results with the von Neumann approach in
Refs.~\onlinecite{lindblad_non-equilibrium_1983,peres_quantum_2002,esposito_entropy_2010}.
One can see that the lack of energy resolution in the von Neumann approach results in appearance of
negative entropy production rate. As a result, integrating over
part of the thermodynamic process may yield 
negative entropy production which contradicts the second law of thermodynamics.
Note that our inside-outside approach yields positive entropy production for an arbitrary initial state.

\begin{figure}[t]
\centering\includegraphics[width=\linewidth]{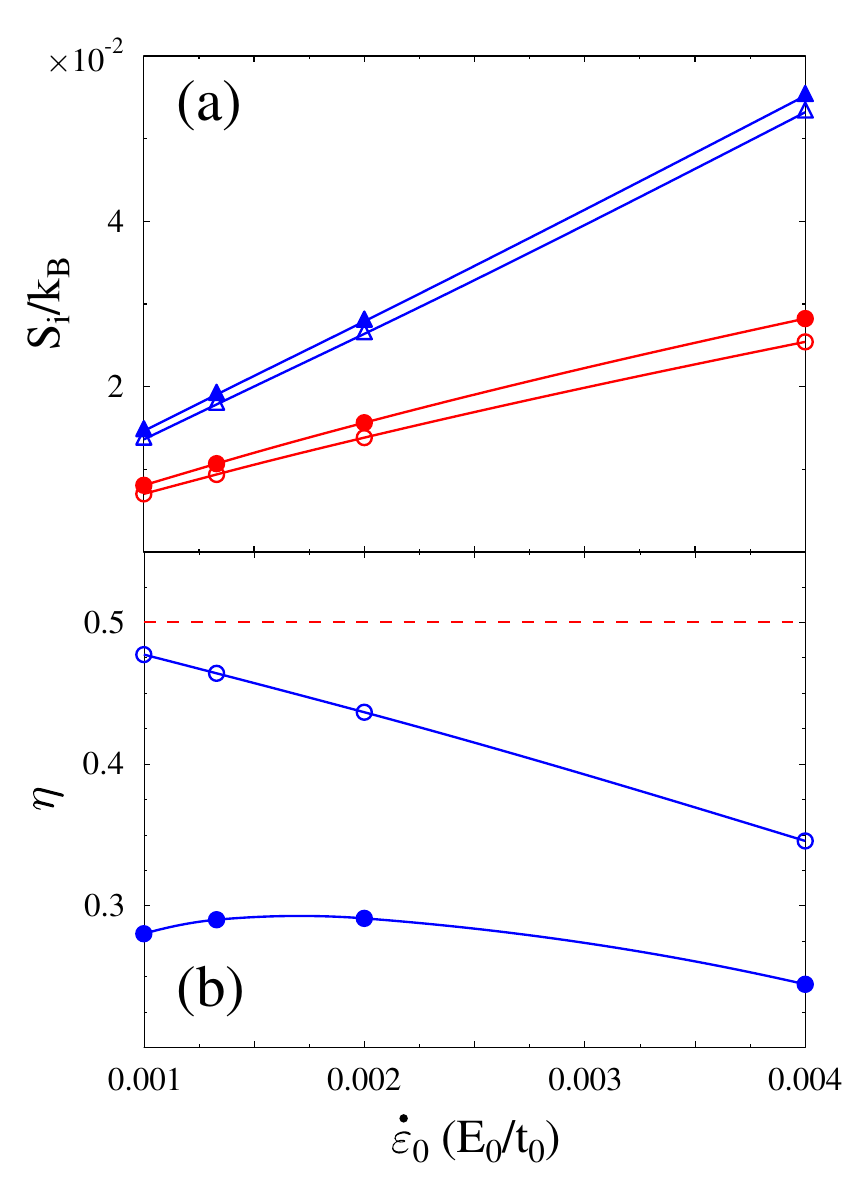}
\caption{\label{fig3}
(Color online) The Carnot cycle for the resonant junction model with (filled markers) 
and without (empty markers) intra-system interaction. Shown are (a) entropy production 
during hot (red line, circles) and cold (blue line, triangles) isothermal parts of the cycle
and (b) efficiency (blue line, circles) of the Carnot cycle vs. driving rate. 
Dashed red line shows the Carnot efficiency of the cycle. 
See text for parameters.
}
\end{figure}

We now consider the Carnot cycle in the resonant junction model 
(see Fig.~\ref{fig1} for a sketch). In step 1 (isothermal part of the cycle with a constant coupling
strength to the hot reservoir) the resonant level is driven from $0.2$ to $0.1$ with a variety of driving rates.
For simplicity, step 2 (decoupling from the hot and subsequent coupling to the cold reservoir) is
performed adiabatically slow. This allows to find an analytical connection between
rates $\dot{u}$ and $\dot{\varepsilon}_0$ from the requirement that $\dot{Q}=0$ during the process
(see Appendix~\ref{appF} for details). Also, this guarantees zero entropy production
during the step. Step 3 is performed under the same rate $\dot{\varepsilon}_0$ as step 1.
Finally, step 4 is again performed adiabatically slow.
Temperature of the hot reservoir is $T_H=0.1$, and the cold reservoir has temperature $T_C=0.05$. 
Figure~\ref{fig3} shows results for the Carnot cycle vs. resonant level driving rate
with (filled markers) and without (empty markers) electron-phonon interaction.
Panel (a) shows entropy productions during the isothermal parts of the cycle. 
As anticipated, entropy production grows with driving rate, and the entropy production for step 3 is 
higher than that of step 1. Also, electron-phonon interaction
increases entropy production due to the presence of additional (bosonic) bath.
Panel (b) shows that the efficiency of the cycle is expectedly lower for faster driving and 
in the presence of interaction.
Note that its dependence on the driving rate is non-monotonic
which is expected for the employed model
(see Ref.~\onlinecite{migliore_relationship_2012} for discussion on the
relevance of the model to phonon assisted electron transport in junctions). 
Indeed, within the model, strength of electron-phonon interaction depends 
on the level population which at intermediate rates drops fast during step 3 of the cycle
thus limiting the amount of heat which can be transferred to the cold bosonic bath.
At even higher rates, the amount of heat from the hot bosonic baths also becomes affected
which eventually leads to decreasing efficiency.


{\bf Conclusion.}
We established a connection between the inside-outside approach originally suggested in
Ref.~\onlinecite{bruch_landauer-buttiker_2018} and the nonequilibrum
Green's function formulation which allowed us to extend the method to interacting systems
with arbitrary drivings in the system and system-baths couplings. 
This generalized thermodynamic formulation applies to 
strong system-bath coupling and is consistent
with the underlying microscopic dynamics (quantum transport).
It overcomes limitations of the only other consistent thermodynamic formulation~\cite{esposito_entropy_2010} 
by satisfying any form of the second law of thermodynamics
for any initial state of a thermodynamic process and for any driving protocol.


\begin{acknowledgments}
We thank Felix von Oppen for many helpful discussions and scientific guidance.
This material is based upon work supported by the National Science Foundation under Grant No. CHE-2154323.
\end{acknowledgments}

\appendix
\begin{widetext}
\section{Derivation of the outgoing flow, Eq.(10)}\label{appA}
In quantum transport particle flux is defined as  the rate of population change 
in the bath~[25]
\begin{equation}
\label{SM_IBN}
 I_B^N(t) = -\sum_{\alpha\in B}\sum_{k\in\alpha}\frac{d}{dt}
 \left\langle \hat c^\dagger_{\alpha k}(t)\hat c_{\alpha k}(t)\right\rangle
 \equiv \sum_{\alpha\in B} I_{\alpha\alpha}(t)
\end{equation}
To account for system-induced coherences between states of the bath
we consider coherence flow. Similar to (\ref{SM_IBN}) it is defined as the rate 
of coherence change
\begin{equation}
 I_{\alpha\beta}(t) = -\frac{d}{dt}\sum_{k\in\alpha}\sum_{k'\in\beta}
 \left\langle \hat c_{\beta k'}^\dagger(t)\,\hat c_{\alpha k}(t)\right\rangle
\end{equation}  
For Hamiltonian (1)-(2) it can be expressed in terms of bath and mixed
Green's functions
\begin{equation}
\begin{split}
 I_{\alpha\beta}(t) &= \sum_{m\in S}\sum_{k\in \alpha}\sum_{k'\in \beta}
 \left(V_{\alpha k,m}(t)\, G^{<}_{m,\beta k'}(t,t)-G^{<}_{\alpha k,m}(t,t)\, V_{m,\beta k'}(t)\right)
 \\ &
 +\sum_{k\in\alpha}\sum_{k'\in\beta}\left(\varepsilon_{\alpha k}-\varepsilon_{\beta k'}\right)
 G^{<}_{\alpha k,\beta k'}(t,t)
\end{split}
\end{equation}
where the Green's functions are the lesser projections of 
\begin{equation}
\begin{split}
 G_{m,\beta k'}(\tau,\tau') &= -i\left\langle T_c\,\hat d_m(\tau)\,\hat c^\dagger_{\beta k'}(\tau')\right\rangle
 \\
 G_{\alpha k,m}(\tau,\tau') &= -i\left\langle T_c\, \hat c_{\alpha k}(\tau)\,\hat d_m^\dagger(\tau')\right\rangle
 \\
 G_{\alpha k,\beta k'}(\tau,\tau') &= -i\left\langle T_c\, \hat c_{\alpha k}(\tau)\, \hat c_{\beta k'}^\dagger(\tau')\right\rangle
\end{split}
\end{equation}
Here, $T_c$ is the Keldysh contour ordering operator, 
and $\tau$ and $\tau'$ are the contour variables.

Using the Dyson equation mixed Green's functions are expressed in terms of free bath 
and system evolutions as
\begin{equation}
\label{SM_Gmix}
\begin{split}
 G_{m,\beta k'}(\tau,\tau') &= \sum_{m_1\in S}\int_c d\tau_1\,
 G_{mm_1}(\tau,\tau_1)\, V_{m_1,\beta k'}(t_1)\,
 g_{\beta k'}(\tau_1,\tau')
 \\
 G_{\alpha k,m}(\tau,\tau') &= \sum_{m_1\in S}\int_c d\tau_1\,
 g_{\alpha k}(\tau,\tau_1)\, V_{\alpha k,m_1}(t_1)\,
 G_{m_1m}(\tau_1,\tau')
\end{split}
\end{equation}
Here, 
\begin{equation}
 g_\alpha(\tau,\tau')\equiv -i \left\langle T_c\, \hat c_{\alpha k}(\tau)\,\hat c^\dagger_{\alpha k}(\tau')\right\rangle_0
\end{equation}
is the Green's function of free evolution in bath channel $\alpha$, and 
system Green's function $G_{m_1m_2}(\tau_1,\tau_2)$ is defined in Eq.(3).

Taking lesser projection of (\ref{SM_Gmix}) and going from sum over states to integrals
over energies
\begin{equation}
 \sum_{k\in \alpha}\ldots = \int\frac{dE_\alpha}{2\pi}\, 2\pi\sum_{k\in\alpha}\delta(E-\varepsilon_{\alpha k})\ldots \equiv \int\frac{dE_\alpha}{2\pi}\, 2\pi\rho_\alpha(E_\alpha)\ldots
\end{equation}
leads to the following expression for the coherence flow
\begin{equation}
\label{SM_Iab}
\begin{split}
I_{\alpha\beta}(t) &= \int\frac{dE_\alpha}{2\pi}\int\frac{dE_\beta}{2\pi}\,
(2\pi)^2\,\rho_\alpha(E_\alpha)\rho_\beta(E_\beta)
\\ &\bigg\{ i\sum_{m,m_1\in S}\int_{-\infty}^{t} dt_1\, \bigg(
e^{-iE_\beta(t_1-t)}\, V_{\alpha,m}(E_\alpha,t)\, V_{m_1,\beta}(E_\beta,t_1)
\left[ G^{<}_{mm_1}(t,t_1)+G^r_{mm_1}(t,t_1)\, f_\beta(E_\beta)\right]
\\ &\qquad\qquad\qquad\qquad + e^{+iE_\alpha (t-t_1)}\, V_{\alpha,m_1}(E_\alpha,t_1)\, V_{m,\beta}(E_\beta,t)
 \left[G^{<}_{m_1m}(t_1,t)-G^a_{m_1m}(t_1,t)\, f_\alpha(E_\alpha)\right]
\bigg)
\\ & + (E_\alpha-E_\beta) G^{<}_{E_\alpha,E_\beta}(t,t)
\bigg\}
\end{split}
\end{equation}
Representing the flow as the difference between incoming and outgoing fluxes
\begin{equation}
I_{\alpha\beta}(t) = \int\frac{dE_\alpha}{2\pi}\int\frac{dE_\beta}{2\pi}
\bigg[\phi_{\alpha\beta}^{in}(E_\alpha,E_\beta)-\phi_{\alpha\beta}^{out}(t;E_\alpha,E_\beta)\bigg]
\end{equation}
and comparing with (\ref{SM_Iab}) leads to the expression presented in Eq.(10).


\section{Derivation of the differential form of the second law, Eq.(7)}\label{appB}
Unitarity of the overall evolution allows to express system entropy change in terms of baths entropy flux
\begin{equation}
\label{SM_dS}
\frac{d}{dt}S(t)=-\frac{d}{dt}S_{B,tot}(t)=
\mbox{Tr}_{c,E}\left\{\sigma\left[\phi^{in}\right]-\sigma\left[\phi^{out}(t)\right]\right\}
\end{equation}
where we used Eq.(11).
According to Eq.(10)
\begin{equation}
\phi^{out}(t)=\phi^{in}+\delta\phi(t)
\end{equation}
Thus, using Eq.(12) we can write
\begin{equation}
\label{SM_sigout}
\begin{split}
& \mbox{Tr}_{c,E}\left\{\sigma\left[\phi^{out}\right]\right\} \equiv \mbox{Tr}_{c,E}\left\{-\phi^{out}\,\ln\phi^{out} 
- \left(\mathbf{I}-\phi^{out}\right)\,\ln\left(\mathbf{I}-\phi^{out}\right)\right\}
\\ &=
 \mbox{Tr}_{c,E}\left\{-\phi^{in}\,\ln\phi^{in} 
- \left(\mathbf{I}-\phi^{in}\right)\,\ln\left(\mathbf{I}-\phi^{in}\right)\right\}
-\mbox{Tr}_{c,E}\left\{\delta\phi\left[\ln\phi^{in}-\ln\left(\mathbf{I}-\phi^{in}\right)\right]\right\}
\\ &+\mbox{Tr}_{c,E}\left\{
-\phi^{out}\left[\ln\phi^{out}-\ln\phi^{in}\right] - \left(\mathbf{I}-\phi^{out}\right)\,
\left[\ln\left(\mathbf{I}-\phi^{out}\right)-\ln\left(\mathbf{I}-\phi^{in}\right)\right]
\right\}
\\ &\equiv
\mbox{Tr}_{c,E}\left\{\sigma\left[\phi^{in}\right]\right\}  - \sum_B \beta_B\dot{Q}_B(t)
- \dot{S}_i(t)
\end{split}
\end{equation}
where to write the last line we used
\begin{equation}
\begin{split}
& \left[\ln\frac{1-\phi^{in}}{\phi^{in}}\right]_{\alpha\beta}
=\delta_{\alpha,\beta}\,\delta(E_\alpha-E_\beta)\,\ln\frac{1-f_\alpha(E_\alpha)}{f_\alpha(E_\alpha)}
\equiv\delta_{\alpha,\beta}\,\delta(E_\alpha-E_\beta)\,\frac{E-\mu_\alpha}{k_BT_\alpha},
\\ &
 \left[\delta\phi(t;E,E)\right]_{\alpha\alpha} = -i_\alpha(t,E),
\end{split}
\end{equation}
and the definition of entropy production rate $\dot{S}_i(t)$, Eq.(13) of the main text.
Substituting (\ref{SM_sigout}) into (\ref{SM_dS}) leads to Eq.(7).


\section{Simplified expressions for coherence flows}\label{appC}
To facilitate numerical simulations we simplify the general expression for outgoing coherence 
flow $\phi^{out}_{\alpha\beta}(t;E_\alpha,E_\beta)$, Eq.(10),
by employing two approximations: 
\begin{enumerate}
\item Wide-band approximation (WBA) in which system-bath coupling is assumed to be
energy independent
\begin{equation}
V_{\alpha,m}(E_\alpha,t)=V_{\alpha,m}(t),\qquad
V_{m,\beta}(E_\beta,t)=V_{m,\beta}(t)
\end{equation}
\item Diagonal approximation  in which the highly oscillating terms are neglected.
Because 
\begin{equation}
 G^{<}_{E_\alpha,E_\beta}(t,t)\sim e^{-i(E_\alpha-E_\beta)t}
\end{equation}
the last term in expression for $\phi^{out}_{\alpha\beta}(t;E_\alpha,E_\beta)$,
Eq.(10), can be dropped because for similar energies  prefactor 
$E_\alpha-E_\beta\sim 0$ and for large differences it is highly oscillating
in time contribution which self-averages to zero.
\end{enumerate}

Within the two approximations the general expression for coherence flow
\begin{equation}
I_{\alpha\beta}(t) =\int\frac{dE_\alpha}{2\pi}\int\frac{dE_\beta}{2\pi}\,
\left[\phi_{\alpha\beta}^{in}(E_\alpha,E_\beta)-\phi_{\alpha\beta}^{out}(t;E_\alpha,E_\beta)\right]
\end{equation}
allows evaluation of one of the integrals and thus reduces the expression to
\begin{equation}
I_{\alpha\beta}(t) = \int\frac{dE}{2\pi}\,
\left[\phi_{\alpha\beta}^{in}(E)-\phi_{\alpha\beta}^{out}(t,E)\right]
\end{equation}
where
\begin{equation}
\begin{split}
\phi_{\alpha\beta}^{in}(E) &= \delta_{\alpha,\beta}\, f_\alpha(E)
\\
\phi_{\alpha\beta}^{out}(t,E) &= \phi_{\alpha\beta}^{in}(E) 
-2\pi\, i \sum_{m,m_1\in S}\int_{-\infty}^{t}dt_1\, 
\\ & \bigg(
e^{-iE(t_1-t)}\, V_{\alpha,m}(t)\, V_{m_1,\beta}(t_1)\,\rho_\beta(E)
\\ &\quad\times
\left[ G^{<}_{mm_1}(t,t_1)+G^{r}_{mm_1}(t,t_1)\, f_\beta(E) \right]
\\ &+
e^{+iE(t_1-t)}\, V_{\alpha,m_1}(t_1)\, V_{m,\beta}(t)\,\rho_\alpha(E)
\\ &\quad\times
\left[ G^{<}_{m_1m}(t_1,t)-G^{a}_{m_1m}(t_1,t)\, f_\alpha(E) \right]
\bigg)
\end{split}
\end{equation}
In this form expression for coherence flow is diagonal in energy.

Comparison with Eqs.~(4) and (5) for $\alpha=\beta$ yields
\begin{equation}
\phi_{\alpha\alpha}^{in}(E) -\phi_{\alpha\alpha}^{out}(t,E)
\equiv f_\alpha(E) - \phi_{\alpha\alpha}^{out}(t,E)
=i_\alpha(t,E)
\end{equation}


\section{Thermodynamic formulation for bosonic baths}\label{appD}
Building the inside-outside approach to system thermodynamics for bosonic baths
is identical to that presented in the main text for the case of fermionic baths.

Expressions for energy-resolved particle flux (analog of Eq.(5) in the main text)
\begin{equation}
\label{SM_iph}
i_\alpha(t,E) \equiv 2\, \mbox{Im}\sum_{v,v_1\in S}\int_{-\infty}^t dt_1\, e^{-iE(t_1-t)}\,
2\pi\rho_\alpha(E)\, U_{v_1,\alpha}(E,t_1)\, U_{\alpha,v}(E,t)
\bigg(F_{vv_1}^{<}(t,t_1)-N_\alpha(E)\, F_{vv_1}^{r}(t,t_1)\bigg),
\end{equation}
incoming and outgoing fluxes (analog of Eq.(10) in the main text)
\begin{equation}
\begin{split}
& \phi_{\alpha\beta}^{in}(E_\alpha,E_\beta) \equiv 2\pi\delta_{\alpha,\beta}\delta\left(E_\alpha-E_\beta\right) N_\alpha(E_\alpha)
\\
& \phi_{\alpha\beta}^{out}(t;E_\alpha,E_\beta) \equiv 
\phi_{\alpha\beta}^{in}(E_\alpha,E_\beta)
-i(2\pi)^2\rho_\alpha(E_\alpha)\rho_\beta(E_\beta)\sum_{v,v_1\in S}\int_{-\infty}^{t} dt_1\,
\\ &
\bigg(\,\,\,
e^{-iE_\beta(t_1-t)}\, U_{\alpha,v}(E_\alpha,t)\, U_{v_1,\beta}(E_\beta,t_1)
\left[ F^{<}_{vv_1}(t,t_1)-F^{r}_{vv_1}(t,t_1)\, N_\beta(E_\beta)\right]
\\ &  +
e^{+iE_\alpha(t-t_1)}\, U_{\alpha,v_1}(E_\alpha,t_1)\, U_{v,\beta}(E_\beta,t)
\left[ F^{<}_{v_1v}(t_1,t)+F^{a}_{v_1v}(t_1,t)\, N_\alpha(E_\alpha)\right]
\bigg)
\\ & - (2\pi)^2 \rho_\alpha(E_\alpha)\rho_\beta(E_\beta) (E_\alpha-E_\beta) 
F^{<}_{E_\alpha,E_\beta}(t,t),
\end{split}
\end{equation}
and entropy production rate (analog of Eq.(13) in the main text)
\begin{equation}
 \dot{S}_i(t) \equiv 
\mbox{Tr}_{c,E}\bigg\{\phi^{out}(t)\,\big[\ln\phi^{out}(t)-\ln\phi^{in}\big]
-\left(\mathbf{I}+\phi^{out}(t)\right)\,\big[\ln\left(\mathbf{I}+\phi^{out}(t)\right)-\ln\left(\mathbf{I}+\phi^{in}\right)\big]
\bigg\}
\end{equation}
are slightly different for bosonic baths.
Here, $v$ indicates molecular vibrational modes, 
$N_\alpha(E)$ is the Bose-Einstein thermal distribution in bath channel $\alpha$,  
and
\begin{equation}
\label{SM_F}
F_{v_1v_2}(\tau_1,\tau_2) \equiv -i\left\langle T_c\, \hat a_{v_1}(\tau_1)\, 
\hat a_{v_2}^\dagger(\tau_2)\right\rangle
\end{equation}
is the single-particle Green's function of molecular vibrations
(compare with Eq.(3) in the main text).

Note that the system does not induce correlations between fermionic and bosonic baths.
That is, contributions of the baths to the thermodynamic formulation are additive. 


\section{Details of numerical simulations}\label{appE}
Central to the inside-outside thermodynamic formulation is the ability to simulate energy-resolved particle fluxes $i_\alpha(t,E)$, Eqs.~(5) and (\ref{SM_iph}). 
For the Holstein model, Eqs.~(14) and (15), and simultaneous coupling 
to one fermionic and one bosonic reservoirs, only one channel $\alpha$ in each bath
is considered. Thus, we will drop the channel index.

To simulate the energy-resolved particle fluxes, we consider the
retarded, lesser and greater projections of the single-particle Green's functions, 
Eqs.~(3) and (\ref{SM_F}), within the wide band approximation. 

In the absence of electron-phonon coupling, $M=0$,
zero-order (in the coupling) Green's functions for the model (14) and (15) are
\begin{equation}
\begin{split}
G_0^r(t_1,t_2) &= -i\theta(t_1-t_2)\, \exp\left(-i\int_{t_2}^{t_1} ds\, \left[\varepsilon_0(s)-\frac{i}{2} u^2(s) \Gamma_0\right]\right)
\\
G_0^{\gtrless}(t_1,t_2) &= \int dt_3\int dt_4\, G_0^r(t_1,t_3)\, u(t_3)\,\Sigma^{\gtrless}_K(t_3,t_4)\,  u(t_4)\,G_0^a(t_4,t_2)
\\
F_0^r(t_1,t_2) &= -i\theta(t_1-t_2)\, \exp\left(-i[\omega_0-i\gamma_0/2][t_1-t_2]\right)
\\
F_0^{\gtrless}(t_1,t_2) &= \int dt_3\int dt_4\, F_0^r(t_1,t_3)\,\Pi^{\gtrless}_P(t_3,t_4)\, F_0^a(t_4,t_2)
\end{split}
\end{equation}
Here, $\theta(t_1-t_2)$ is the Heaviside step function, $\Sigma_K$ and $\Pi_P$
are respectively the electron self-energy due to coupling to the fermionic bath and vibration
self-energy due to coupling to the thermal bath. The latter are Fourier transforms of
\begin{equation}
\begin{split}
\Sigma_K^{>}(E) &= -i\Gamma_0\left[1-f(E)\right],\qquad
\Sigma_K^{<}(E)=+i\Gamma_0 f(E),
\\
\Pi_P^{>}(E) &= -i\gamma_0\left[1+N(E)\right],\qquad
\Pi_P^{<}(E)=-i\gamma_0 N(E).
\end{split}
\end{equation}

In the presence of electron-phonon interaction, Green's functions should
be obtained within a self-consistent procedure using system of coupled 
Dyson equations
\begin{equation}
\begin{split}
G^{\gtrless}(t_1,t_2) &= \int dt_3\int dt_4\, G_0^r(t_1,t_3)\left[ 
u(t_3)\,\Sigma^{\gtrless}_K(t_3,t_4)\, u(t_4)+\Sigma^{\gtrless}_v(t_3,t_4)\right]G^a(t_4,t_2)
\\
F^{\gtrless}(t_1,t_2) &= \int dt_3\int dt_4\, F^r(t_1,t_3)\left[\Pi^{\gtrless}_P(t_3,t_4)+\Pi^{\gtrless}_e(t_3,t_4)\right] F^a(t_4,t_2)
\end{split}
\end{equation}
where $G^r(t_1,t_2)\equiv\theta(t_1-t_2)\left[G^{>}(t_1,t_2)-G^{<}(t_1,t_2)\right]$,
$F^r(t_1,t_2)\equiv\theta(t_1-t_2)\left[G^{>}(t_1,t_2)-G^{<}(t_1,t_2)\right]$,
$G^a(t_1,t_2)\equiv \left[G^r(t_2,t_1)\right]^{*}$ and
$F^a(t_1,t_2)\equiv \left[F^r(t_2,t_1)\right]^{*}$.
 $\Sigma_v$ and $\Pi_e$ are respectively the electron self-energy due to 
interaction with vibration
and the vibration self-energy due to interaction with electron.
Within the SCBA they are
\begin{equation}
\begin{split}
\Sigma_v(\tau_1,\tau_2) &=+i\, M^2\, G(\tau_1,\tau_2)\left[F(\tau_1,\tau_2)+F(\tau_2,\tau_1)\right]
\\
\Pi_e(\tau_1,\tau_2) &= -i\, M^2\, G(\tau_1,\tau_2)\, G(\tau_2,\tau_1)
\end{split}
\end{equation}
Here, in the electron self-energy we neglected the Hartree term.

Once Green's functions are available, we dress electron Green's function with 
the system-bath coupling rate $u(t)$,
\begin{equation}
 D(\tau_1,\tau_2)\equiv u(t_1)\, G(\tau_1,\tau_2)\, u(t_2)
\end{equation}
consider retarded parts of the lesser and greater projections,
\begin{equation}
 D^{\gtrless\, +}(t_1,t_2)\equiv \theta(t_1-t_2)D^{\gtrless}(t_1,t_2),\qquad
 F^{\gtrless\, +}(t_1,t_2)\equiv \theta(t_1-t_2)F^{\gtrless}(t_1,t_2),
\end{equation} 
and perform one-sided Fourier transform
\begin{equation}
 D^{\gtrless\, +}(t,E) \equiv \int_{-\infty}^{+\infty} dt_1\, e^{-iE t_1}\, D^{\gtrless\, +}(t,t_1),\qquad
 F^{\gtrless\, +}(t,E) \equiv \int_{-\infty}^{+\infty} dt_1\, e^{-iE t_1}\, F^{\gtrless\, +}(t,t_1)
\end{equation}
In terms of these Fourier transforms electron and vibration energy-resolved particle fluxes,
Eqs.~(5) and (\ref{SM_iph}), are
\begin{equation}
\label{SM_itE}
\begin{split}
 i_e(t,E) &=-2\,\mbox{Im}\left\{
 e^{iEt}\,\Gamma(E)\left(f\,(E) D^{>\, +}(t,E)+[1-f\,(E)] D^{<\, +}(t,E) \right)
 \right\}
\\
i_v(t,E) &=-2\,\mbox{Im}\left\{
 e^{iEt}\,\gamma(E)\left(N(E) F^{>\, +}(t,E) - \left[1+N(E)\right] F^{<\, +}(t,E)\right)
 \right\}
 \end{split}
\end{equation}
Here, $\Gamma(E)$ and $\gamma(E)$ are normalized values of escape rates.
Eq.(\ref{SM_itE}) is used to calculate thermodynamic properties as indicated in the main text
and the section above.


\section{Connection between the level and coupling driving rates during adiabatic (de)coupling process}\label{appF}
In the absence of electron-vibration coupling, $M=0$, expressions for adiabatic driving (and beyond) were derived in Ref.~15. Specifically, heat flux under
adiabatic driving is given in Eq.(10) of that paper. In accordance with the standard 
quantum transport definitions we take $\alpha=0$ in this expression
\begin{equation}
\label{SM_dotQ}
\begin{split}
\dot{Q}^{(1)}(t) &=\frac{d}{dt}\int\frac{dE}{2\pi}\, f(E) A^{(0)}(t,E)\left(2E-\varepsilon_0(t)-\mu\right)
\\ &
-\int\frac{dE}{2\pi} f(E)\left(A^{(0)}(t,E)\left[\dot{\varepsilon}_0+\dot{\Lambda}(t,E)\right]+
\mbox{Re}\, G^{r\, (0)}(t,E)\,\dot{\Gamma}(t,E)\right)
\end{split}
\end{equation}
where 
\begin{equation}
\begin{split}
 A^{(0)}(t,E) &\equiv \frac{\Gamma(t,E)}{\left[E-\varepsilon_0(t)-\Lambda(t,E)\right]^2+\left[\Gamma(t,E)/2\right]^2}
 \\
 \mbox{Re}\, G^{r\, (0)}(t,E) &\equiv \frac{E-\varepsilon_0(t)-\Lambda(t,E)}{\left[E-\varepsilon_0(t)-\Lambda(t,E)\right]^2+\left[\Gamma(t,E)/2\right]^2}
\end{split}
\end{equation}
are the zero order (in driving) expressions for the spectral function and real part of the retarded projection
of the Green's function (3), and the Lamb shift $\Lambda(t,E)$ and the broadening 
$\Gamma(t,E)$ 
are~[15] 
\begin{equation}
\begin{split}
\Lambda(t,E) &= u^2(t)\frac{\Gamma_0}{2}\frac{(E-E_B)W_B}{(E-E_B)^2+W_B^2}
\\
\Gamma(t,E) &= u^2(t)\Gamma_0\frac{W_B^2}{(E-E_B)^2+W_B^2}
\end{split}
\end{equation}
where $E_B$ and $W_B$ are the center and width of the band, respectively, and
$\Gamma_0$ is the level escape rate.

Adiabatic coupling/decoupling is defined by the $\dot{Q}^{(1)}(t)=0$ condition. Using (\ref{SM_dotQ})
and employing
\begin{equation}
\begin{split}
\dot{\Lambda}(t,E) &= 2\frac{\dot{u}}{u(t)}\Lambda(t,E)
\\
\dot{\Gamma}(t,E) &= 2\frac{\dot{u}}{u(t)}\Gamma(t,E)
\end{split}
\end{equation}
leads to the following connection between the driving rates
\begin{equation}
\begin{split}
\dot{\varepsilon}_0 &= \frac{\dot{u}}{u(t)}\, 
\int\frac{dE}{2\pi}\, f(E)\left([2E-\varepsilon_0(t)-\mu][1+\Lambda(t,E)(2\,\mbox{Re}\, G^{r\, (0)}(t,E)-1)-\Gamma(t,E)\, A^{(0)}(t,E)/2]\,  A^{(0)}(t,E) \right.
\\ & \left.\qquad\qquad\qquad\qquad
- \Gamma(t,E)\,\mbox{Re}\, G^{r\, (0)}(t,E) \right)
\bigg/
\int\frac{dE}{2\pi}\, f(E) A^{(0)}(t,E)\left(1-[2E-\varepsilon_0(t)-\mu]\,\mbox{Re}\, G^{r\, (0)}(t,E)\right)
\end{split}
\end{equation}

In the presence of electron-vibration coupling, we substitute $A^{(0)}$ and $G^{r\, (0)}$
with their SCBA analogs. This is an approximation.
\end{widetext}


\begin{thebibliography}{38}
\expandafter\ifx\csname natexlab\endcsname\relax\def\natexlab#1{#1}\fi
\expandafter\ifx\csname bibnamefont\endcsname\relax
  \def\bibnamefont#1{#1}\fi
\expandafter\ifx\csname bibfnamefont\endcsname\relax
  \def\bibfnamefont#1{#1}\fi
\expandafter\ifx\csname citenamefont\endcsname\relax
  \def\citenamefont#1{#1}\fi
\expandafter\ifx\csname url\endcsname\relax
  \def\url#1{\texttt{#1}}\fi
\expandafter\ifx\csname urlprefix\endcsname\relax\def\urlprefix{URL }\fi
\providecommand{\bibinfo}[2]{#2}
\providecommand{\eprint}[2][]{\url{#2}}

\bibitem[{\citenamefont{Jezouin et~al.}(2013)\citenamefont{Jezouin, Parmentier,
  Anthore, Gennser, Cavanna, Jin, and Pierre}}]{jezouin_quantum_2013}
\bibinfo{author}{\bibfnamefont{S.}~\bibnamefont{Jezouin}},
  \bibinfo{author}{\bibfnamefont{F.~D.} \bibnamefont{Parmentier}},
  \bibinfo{author}{\bibfnamefont{A.}~\bibnamefont{Anthore}},
  \bibinfo{author}{\bibfnamefont{U.}~\bibnamefont{Gennser}},
  \bibinfo{author}{\bibfnamefont{A.}~\bibnamefont{Cavanna}},
  \bibinfo{author}{\bibfnamefont{Y.}~\bibnamefont{Jin}}, \bibnamefont{and}
  \bibinfo{author}{\bibfnamefont{F.}~\bibnamefont{Pierre}},
  \bibinfo{journal}{Science} \textbf{\bibinfo{volume}{342}},
  \bibinfo{pages}{601} (\bibinfo{year}{2013}).

\bibitem[{\citenamefont{Pekola}(2015)}]{pekola_towards_2015}
\bibinfo{author}{\bibfnamefont{J.~P.} \bibnamefont{Pekola}},
  \bibinfo{journal}{Nature Phys.} \textbf{\bibinfo{volume}{11}},
  \bibinfo{pages}{118} (\bibinfo{year}{2015}).

\bibitem[{\citenamefont{Hartman et~al.}(2018)\citenamefont{Hartman, Olsen, L{\"
  u}scher, Samani, Fallahi, Gardner, Manfra, and Folk}}]{hartman_direct_2018}
\bibinfo{author}{\bibfnamefont{N.}~\bibnamefont{Hartman}},
  \bibinfo{author}{\bibfnamefont{C.}~\bibnamefont{Olsen}},
  \bibinfo{author}{\bibfnamefont{S.}~\bibnamefont{L{\" u}scher}},
  \bibinfo{author}{\bibfnamefont{M.}~\bibnamefont{Samani}},
  \bibinfo{author}{\bibfnamefont{S.}~\bibnamefont{Fallahi}},
  \bibinfo{author}{\bibfnamefont{G.~C.} \bibnamefont{Gardner}},
  \bibinfo{author}{\bibfnamefont{M.}~\bibnamefont{Manfra}}, \bibnamefont{and}
  \bibinfo{author}{\bibfnamefont{J.}~\bibnamefont{Folk}},
  \bibinfo{journal}{Nature Phys.} \textbf{\bibinfo{volume}{14}},
  \bibinfo{pages}{1083} (\bibinfo{year}{2018}).

\bibitem[{\citenamefont{Klatzow et~al.}(2019)\citenamefont{Klatzow, Becker,
  Ledingham, Weinzetl, Kaczmarek, Saunders, Nunn, Walmsley, Uzdin, and
  Poem}}]{klatzow_experimental_2019}
\bibinfo{author}{\bibfnamefont{J.}~\bibnamefont{Klatzow}},
  \bibinfo{author}{\bibfnamefont{J.~N.} \bibnamefont{Becker}},
  \bibinfo{author}{\bibfnamefont{P.~M.} \bibnamefont{Ledingham}},
  \bibinfo{author}{\bibfnamefont{C.}~\bibnamefont{Weinzetl}},
  \bibinfo{author}{\bibfnamefont{K.~T.} \bibnamefont{Kaczmarek}},
  \bibinfo{author}{\bibfnamefont{D.~J.} \bibnamefont{Saunders}},
  \bibinfo{author}{\bibfnamefont{J.}~\bibnamefont{Nunn}},
  \bibinfo{author}{\bibfnamefont{I.~A.} \bibnamefont{Walmsley}},
  \bibinfo{author}{\bibfnamefont{R.}~\bibnamefont{Uzdin}}, \bibnamefont{and}
  \bibinfo{author}{\bibfnamefont{E.}~\bibnamefont{Poem}},
  \bibinfo{journal}{Phys. Rev. Lett.} \textbf{\bibinfo{volume}{122}},
  \bibinfo{pages}{110601} (\bibinfo{year}{2019}).

\bibitem[{\citenamefont{Reddy et~al.}(2007)\citenamefont{Reddy, Jang, Segalman,
  and Majumdar}}]{reddy_thermoelectricity_2007}
\bibinfo{author}{\bibfnamefont{P.}~\bibnamefont{Reddy}},
  \bibinfo{author}{\bibfnamefont{S.-Y.} \bibnamefont{Jang}},
  \bibinfo{author}{\bibfnamefont{R.~A.} \bibnamefont{Segalman}},
  \bibnamefont{and} \bibinfo{author}{\bibfnamefont{A.}~\bibnamefont{Majumdar}},
  \bibinfo{journal}{Science} \textbf{\bibinfo{volume}{315}},
  \bibinfo{pages}{1568} (\bibinfo{year}{2007}).

\bibitem[{\citenamefont{Lee et~al.}(2013)\citenamefont{Lee, Kim, Jeong, Zotti,
  Pauly, Cuevas, and Reddy}}]{lee_heat_2013}
\bibinfo{author}{\bibfnamefont{W.}~\bibnamefont{Lee}},
  \bibinfo{author}{\bibfnamefont{K.}~\bibnamefont{Kim}},
  \bibinfo{author}{\bibfnamefont{W.}~\bibnamefont{Jeong}},
  \bibinfo{author}{\bibfnamefont{L.~A.} \bibnamefont{Zotti}},
  \bibinfo{author}{\bibfnamefont{F.}~\bibnamefont{Pauly}},
  \bibinfo{author}{\bibfnamefont{J.~C.} \bibnamefont{Cuevas}},
  \bibnamefont{and} \bibinfo{author}{\bibfnamefont{P.}~\bibnamefont{Reddy}},
  \bibinfo{journal}{Nature} \textbf{\bibinfo{volume}{498}},
  \bibinfo{pages}{209} (\bibinfo{year}{2013}).

\bibitem[{\citenamefont{Kim et~al.}(2014)\citenamefont{Kim, Jeong, Kim, Lee,
  and Reddy}}]{kim_electrostatic_2014}
\bibinfo{author}{\bibfnamefont{Y.}~\bibnamefont{Kim}},
  \bibinfo{author}{\bibfnamefont{W.}~\bibnamefont{Jeong}},
  \bibinfo{author}{\bibfnamefont{K.}~\bibnamefont{Kim}},
  \bibinfo{author}{\bibfnamefont{W.}~\bibnamefont{Lee}}, \bibnamefont{and}
  \bibinfo{author}{\bibfnamefont{P.}~\bibnamefont{Reddy}},
  \bibinfo{journal}{Nature Nanotechnol.} \textbf{\bibinfo{volume}{9}},
  \bibinfo{pages}{881} (\bibinfo{year}{2014}).

\bibitem[{\citenamefont{Zotti et~al.}(2014)\citenamefont{Zotti, B{\" u}rkle,
  Pauly, Lee, Kim, Jeong, Asai, Reddy, and Cuevas}}]{zotti_heat_2014}
\bibinfo{author}{\bibfnamefont{L.~A.} \bibnamefont{Zotti}},
  \bibinfo{author}{\bibfnamefont{M.}~\bibnamefont{B{\" u}rkle}},
  \bibinfo{author}{\bibfnamefont{F.}~\bibnamefont{Pauly}},
  \bibinfo{author}{\bibfnamefont{W.}~\bibnamefont{Lee}},
  \bibinfo{author}{\bibfnamefont{K.}~\bibnamefont{Kim}},
  \bibinfo{author}{\bibfnamefont{W.}~\bibnamefont{Jeong}},
  \bibinfo{author}{\bibfnamefont{Y.}~\bibnamefont{Asai}},
  \bibinfo{author}{\bibfnamefont{P.}~\bibnamefont{Reddy}}, \bibnamefont{and}
  \bibinfo{author}{\bibfnamefont{J.~C.} \bibnamefont{Cuevas}},
  \bibinfo{journal}{New J. Phys.} \textbf{\bibinfo{volume}{16}},
  \bibinfo{pages}{015004} (\bibinfo{year}{2014}).

\bibitem[{\citenamefont{Cui et~al.}(2017{\natexlab{a}})\citenamefont{Cui, Miao,
  Jiang, Meyhofer, and Reddy}}]{cui_perspective:_2017}
\bibinfo{author}{\bibfnamefont{L.}~\bibnamefont{Cui}},
  \bibinfo{author}{\bibfnamefont{R.}~\bibnamefont{Miao}},
  \bibinfo{author}{\bibfnamefont{C.}~\bibnamefont{Jiang}},
  \bibinfo{author}{\bibfnamefont{E.}~\bibnamefont{Meyhofer}}, \bibnamefont{and}
  \bibinfo{author}{\bibfnamefont{P.}~\bibnamefont{Reddy}}, \bibinfo{journal}{J.
  Chem. Phys.} \textbf{\bibinfo{volume}{146}}, \bibinfo{pages}{092201}
  (\bibinfo{year}{2017}{\natexlab{a}}).

\bibitem[{\citenamefont{Cui et~al.}(2017{\natexlab{b}})\citenamefont{Cui,
  Jeong, Hur, Matt, Kl{\" o}ckner, Pauly, Nielaba, Cuevas, Meyhofer, and
  Reddy}}]{cui_quantized_2017}
\bibinfo{author}{\bibfnamefont{L.}~\bibnamefont{Cui}},
  \bibinfo{author}{\bibfnamefont{W.}~\bibnamefont{Jeong}},
  \bibinfo{author}{\bibfnamefont{S.}~\bibnamefont{Hur}},
  \bibinfo{author}{\bibfnamefont{M.}~\bibnamefont{Matt}},
  \bibinfo{author}{\bibfnamefont{J.~C.} \bibnamefont{Kl{\" o}ckner}},
  \bibinfo{author}{\bibfnamefont{F.}~\bibnamefont{Pauly}},
  \bibinfo{author}{\bibfnamefont{P.}~\bibnamefont{Nielaba}},
  \bibinfo{author}{\bibfnamefont{J.~C.} \bibnamefont{Cuevas}},
  \bibinfo{author}{\bibfnamefont{E.}~\bibnamefont{Meyhofer}}, \bibnamefont{and}
  \bibinfo{author}{\bibfnamefont{P.}~\bibnamefont{Reddy}},
  \bibinfo{journal}{Science} \textbf{\bibinfo{volume}{355}},
  \bibinfo{pages}{1192} (\bibinfo{year}{2017}{\natexlab{b}}).

\bibitem[{\citenamefont{Cui et~al.}(2018)\citenamefont{Cui, Miao, Wang,
  Thompson, Zotti, Cuevas, Meyhofer, and Reddy}}]{cui_peltier_2018}
\bibinfo{author}{\bibfnamefont{L.}~\bibnamefont{Cui}},
  \bibinfo{author}{\bibfnamefont{R.}~\bibnamefont{Miao}},
  \bibinfo{author}{\bibfnamefont{K.}~\bibnamefont{Wang}},
  \bibinfo{author}{\bibfnamefont{D.}~\bibnamefont{Thompson}},
  \bibinfo{author}{\bibfnamefont{L.~A.} \bibnamefont{Zotti}},
  \bibinfo{author}{\bibfnamefont{J.~C.} \bibnamefont{Cuevas}},
  \bibinfo{author}{\bibfnamefont{E.}~\bibnamefont{Meyhofer}}, \bibnamefont{and}
  \bibinfo{author}{\bibfnamefont{P.}~\bibnamefont{Reddy}},
  \bibinfo{journal}{Nature Nanotechnol.} \textbf{\bibinfo{volume}{13}},
  \bibinfo{pages}{122} (\bibinfo{year}{2018}).

\bibitem[{\citenamefont{Cui et~al.}(2019)\citenamefont{Cui, Hur, Akbar, Kl{\"
  o}ckner, Jeong, Pauly, Jang, Reddy, and Meyhofer}}]{cui_thermal_2019}
\bibinfo{author}{\bibfnamefont{L.}~\bibnamefont{Cui}},
  \bibinfo{author}{\bibfnamefont{S.}~\bibnamefont{Hur}},
  \bibinfo{author}{\bibfnamefont{Z.~A.} \bibnamefont{Akbar}},
  \bibinfo{author}{\bibfnamefont{J.~C.} \bibnamefont{Kl{\" o}ckner}},
  \bibinfo{author}{\bibfnamefont{W.}~\bibnamefont{Jeong}},
  \bibinfo{author}{\bibfnamefont{F.}~\bibnamefont{Pauly}},
  \bibinfo{author}{\bibfnamefont{S.-Y.} \bibnamefont{Jang}},
  \bibinfo{author}{\bibfnamefont{P.}~\bibnamefont{Reddy}}, \bibnamefont{and}
  \bibinfo{author}{\bibfnamefont{E.}~\bibnamefont{Meyhofer}},
  \bibinfo{journal}{Nature} \textbf{\bibinfo{volume}{572}},
  \bibinfo{pages}{628} (\bibinfo{year}{2019}).

\bibitem[{\citenamefont{Esposito et~al.}(2009)\citenamefont{Esposito, Harbola,
  and Mukamel}}]{esposito_nonequilibrium_2009}
\bibinfo{author}{\bibfnamefont{M.}~\bibnamefont{Esposito}},
  \bibinfo{author}{\bibfnamefont{U.}~\bibnamefont{Harbola}}, \bibnamefont{and}
  \bibinfo{author}{\bibfnamefont{S.}~\bibnamefont{Mukamel}},
  \bibinfo{journal}{Rev. Mod. Phys.} \textbf{\bibinfo{volume}{81}},
  \bibinfo{pages}{1665} (\bibinfo{year}{2009}).

\bibitem[{\citenamefont{Esposito et~al.}(2014)\citenamefont{Esposito, Harbola,
  and Mukamel}}]{esposito_erratum:_2014}
\bibinfo{author}{\bibfnamefont{M.}~\bibnamefont{Esposito}},
  \bibinfo{author}{\bibfnamefont{U.}~\bibnamefont{Harbola}}, \bibnamefont{and}
  \bibinfo{author}{\bibfnamefont{S.}~\bibnamefont{Mukamel}},
  \bibinfo{journal}{Rev. Mod. Phys.} \textbf{\bibinfo{volume}{86}},
  \bibinfo{pages}{1125} (\bibinfo{year}{2014}).

\bibitem[{\citenamefont{Esposito
  et~al.}(2015{\natexlab{a}})\citenamefont{Esposito, Ochoa, and
  Galperin}}]{esposito_nature_2015}
\bibinfo{author}{\bibfnamefont{M.}~\bibnamefont{Esposito}},
  \bibinfo{author}{\bibfnamefont{M.~A.} \bibnamefont{Ochoa}}, \bibnamefont{and}
  \bibinfo{author}{\bibfnamefont{M.}~\bibnamefont{Galperin}},
  \bibinfo{journal}{Phys. Rev. B} \textbf{\bibinfo{volume}{92}},
  \bibinfo{pages}{235440} (\bibinfo{year}{2015}{\natexlab{a}}).

\bibitem[{\citenamefont{Brandner et~al.}(2018)\citenamefont{Brandner, Hanazato,
  and Saito}}]{brandner_thermodynamic_2018}
\bibinfo{author}{\bibfnamefont{K.}~\bibnamefont{Brandner}},
  \bibinfo{author}{\bibfnamefont{T.}~\bibnamefont{Hanazato}}, \bibnamefont{and}
  \bibinfo{author}{\bibfnamefont{K.}~\bibnamefont{Saito}},
  \bibinfo{journal}{Phys. Rev. Lett.} \textbf{\bibinfo{volume}{120}},
  \bibinfo{pages}{090601} (\bibinfo{year}{2018}).

\bibitem[{\citenamefont{Kosloff}(2013)}]{kosloff_quantum_2013}
\bibinfo{author}{\bibfnamefont{R.}~\bibnamefont{Kosloff}},
  \bibinfo{journal}{Entropy} \textbf{\bibinfo{volume}{15}},
  \bibinfo{pages}{2100} (\bibinfo{year}{2013}).

\bibitem[{\citenamefont{Lindblad}(1983)}]{lindblad_non-equilibrium_1983}
\bibinfo{author}{\bibfnamefont{G.}~\bibnamefont{Lindblad}},
  \emph{\bibinfo{title}{Non-{Equilibrium} {Entropy} and {Irreversibility}}}
  (\bibinfo{publisher}{D. Reidel Publishing Company}, \bibinfo{year}{1983}).

\bibitem[{\citenamefont{Peres}(2002)}]{peres_quantum_2002}
\bibinfo{author}{\bibfnamefont{A.}~\bibnamefont{Peres}},
  \emph{\bibinfo{title}{Quantum {Theory}: {Concepts} and {Methods}}}
  (\bibinfo{publisher}{Kluwer Academic Publishers}, \bibinfo{year}{2002}).

\bibitem[{\citenamefont{Esposito et~al.}(2010)\citenamefont{Esposito,
  Lindenberg, and Broeck}}]{esposito_entropy_2010}
\bibinfo{author}{\bibfnamefont{M.}~\bibnamefont{Esposito}},
  \bibinfo{author}{\bibfnamefont{K.}~\bibnamefont{Lindenberg}},
  \bibnamefont{and} \bibinfo{author}{\bibfnamefont{C.~V.~d.}
  \bibnamefont{Broeck}}, \bibinfo{journal}{New J. Phys.}
  \textbf{\bibinfo{volume}{12}}, \bibinfo{pages}{013013}
  (\bibinfo{year}{2010}).

\bibitem[{\citenamefont{Sagawa}(2012)}]{sagawa_second_2012}
\bibinfo{author}{\bibfnamefont{T.}~\bibnamefont{Sagawa}},
  \bibinfo{journal}{arXiv:1202.0983 [cond-mat, physics:math-ph,
  physics:quant-ph]} \textbf{\bibinfo{volume}{8}}, \bibinfo{pages}{125}
  (\bibinfo{year}{2012}), \bibinfo{note}{arXiv: 1202.0983}.

\bibitem[{\citenamefont{Kato and Tanimura}(2016)}]{kato_quantum_2016}
\bibinfo{author}{\bibfnamefont{A.}~\bibnamefont{Kato}} \bibnamefont{and}
  \bibinfo{author}{\bibfnamefont{Y.}~\bibnamefont{Tanimura}},
  \bibinfo{journal}{J. Chem. Phys.} \textbf{\bibinfo{volume}{145}},
  \bibinfo{pages}{224105} (\bibinfo{year}{2016}).

\bibitem[{\citenamefont{Strasberg et~al.}(2017)\citenamefont{Strasberg,
  Schaller, Brandes, and Esposito}}]{strasberg_quantum_2017}
\bibinfo{author}{\bibfnamefont{P.}~\bibnamefont{Strasberg}},
  \bibinfo{author}{\bibfnamefont{G.}~\bibnamefont{Schaller}},
  \bibinfo{author}{\bibfnamefont{T.}~\bibnamefont{Brandes}}, \bibnamefont{and}
  \bibinfo{author}{\bibfnamefont{M.}~\bibnamefont{Esposito}},
  \bibinfo{journal}{Phys. Rev. X} \textbf{\bibinfo{volume}{7}}
  (\bibinfo{year}{2017}).

\bibitem[{\citenamefont{Seshadri and Galperin}(2021)}]{seshadri_entropy_2021}
\bibinfo{author}{\bibfnamefont{N.}~\bibnamefont{Seshadri}} \bibnamefont{and}
  \bibinfo{author}{\bibfnamefont{M.}~\bibnamefont{Galperin}},
  \bibinfo{journal}{Phys. Rev. B} \textbf{\bibinfo{volume}{103}},
  \bibinfo{pages}{085415} (\bibinfo{year}{2021}).

\bibitem[{\citenamefont{Bruch et~al.}(2018)\citenamefont{Bruch, Lewenkopf, and
  von Oppen}}]{bruch_landauer-buttiker_2018}
\bibinfo{author}{\bibfnamefont{A.}~\bibnamefont{Bruch}},
  \bibinfo{author}{\bibfnamefont{C.}~\bibnamefont{Lewenkopf}},
  \bibnamefont{and} \bibinfo{author}{\bibfnamefont{F.}~\bibnamefont{von
  Oppen}}, \bibinfo{journal}{Phys. Rev. Lett.} \textbf{\bibinfo{volume}{120}},
  \bibinfo{pages}{107701} (\bibinfo{year}{2018}).

\bibitem[{\citenamefont{Haug and Jauho}(2008)}]{haug_quantum_2008}
\bibinfo{author}{\bibfnamefont{H.}~\bibnamefont{Haug}} \bibnamefont{and}
  \bibinfo{author}{\bibfnamefont{A.-P.} \bibnamefont{Jauho}},
  \emph{\bibinfo{title}{Quantum {Kinetics} in {Transport} and {Optics} of
  {Semiconductors}}} (\bibinfo{publisher}{Springer}, \bibinfo{address}{Berlin
  Heidelberg}, \bibinfo{year}{2008}), \bibinfo{edition}{second, substantially
  revised edition} ed.

\bibitem[{\citenamefont{Stefanucci and van
  Leeuwen}(2013)}]{stefanucci_nonequilibrium_2013}
\bibinfo{author}{\bibfnamefont{G.}~\bibnamefont{Stefanucci}} \bibnamefont{and}
  \bibinfo{author}{\bibfnamefont{R.}~\bibnamefont{van Leeuwen}},
  \emph{\bibinfo{title}{Nonequilibrium {Many}-{Body} {Theory} of {Quantum}
  {Systems}. {A} {Modern} {Introduction}.}} (\bibinfo{publisher}{Cambridge
  University Press}, \bibinfo{year}{2013}).

\bibitem[{\citenamefont{Jauho et~al.}(1994)\citenamefont{Jauho, Wingreen, and
  Meir}}]{jauho_time-dependent_1994}
\bibinfo{author}{\bibfnamefont{A.-P.} \bibnamefont{Jauho}},
  \bibinfo{author}{\bibfnamefont{N.~S.} \bibnamefont{Wingreen}},
  \bibnamefont{and} \bibinfo{author}{\bibfnamefont{Y.}~\bibnamefont{Meir}},
  \bibinfo{journal}{Phys. Rev. B} \textbf{\bibinfo{volume}{50}},
  \bibinfo{pages}{5528} (\bibinfo{year}{1994}).

\bibitem[{\citenamefont{Galperin et~al.}(2007)\citenamefont{Galperin, Nitzan,
  and Ratner}}]{galperin_heat_2007}
\bibinfo{author}{\bibfnamefont{M.}~\bibnamefont{Galperin}},
  \bibinfo{author}{\bibfnamefont{A.}~\bibnamefont{Nitzan}}, \bibnamefont{and}
  \bibinfo{author}{\bibfnamefont{M.~A.} \bibnamefont{Ratner}},
  \bibinfo{journal}{Phys. Rev. B} \textbf{\bibinfo{volume}{75}},
  \bibinfo{pages}{155312} (\bibinfo{year}{2007}).

\bibitem[{\citenamefont{Vedral}(2002)}]{vedral_role_2002}
\bibinfo{author}{\bibfnamefont{V.}~\bibnamefont{Vedral}},
  \bibinfo{journal}{Rev. Mod. Phys.} \textbf{\bibinfo{volume}{74}},
  \bibinfo{pages}{197} (\bibinfo{year}{2002}),
  \urlprefix\url{https://link.aps.org/doi/10.1103/RevModPhys.74.197}.

\bibitem[{\citenamefont{Esposito
  et~al.}(2015{\natexlab{b}})\citenamefont{Esposito, Ochoa, and
  Galperin}}]{esposito_quantum_2015}
\bibinfo{author}{\bibfnamefont{M.}~\bibnamefont{Esposito}},
  \bibinfo{author}{\bibfnamefont{M.~A.} \bibnamefont{Ochoa}}, \bibnamefont{and}
  \bibinfo{author}{\bibfnamefont{M.}~\bibnamefont{Galperin}},
  \bibinfo{journal}{Phys. Rev. Lett.} \textbf{\bibinfo{volume}{114}},
  \bibinfo{pages}{080602} (\bibinfo{year}{2015}{\natexlab{b}}).

\bibitem[{\citenamefont{Park and Galperin}(2011)}]{park_self-consistent_2011}
\bibinfo{author}{\bibfnamefont{T.-H.} \bibnamefont{Park}} \bibnamefont{and}
  \bibinfo{author}{\bibfnamefont{M.}~\bibnamefont{Galperin}},
  \bibinfo{journal}{Phys. Rev. B} \textbf{\bibinfo{volume}{84}},
  \bibinfo{pages}{205450} (\bibinfo{year}{2011}).

\bibitem[{\citenamefont{Frigo and Johnson}(2005)}]{FFTW}
\bibinfo{author}{\bibfnamefont{M.}~\bibnamefont{Frigo}} \bibnamefont{and}
  \bibinfo{author}{\bibfnamefont{S.~G.} \bibnamefont{Johnson}},
  \bibinfo{journal}{Proc. IEEE} \textbf{\bibinfo{volume}{93}},
  \bibinfo{pages}{216} (\bibinfo{year}{2005}).

\bibitem[{\citenamefont{Migliore et~al.}(2012)\citenamefont{Migliore, Schiff,
  and Nitzan}}]{migliore_relationship_2012}
\bibinfo{author}{\bibfnamefont{A.}~\bibnamefont{Migliore}},
  \bibinfo{author}{\bibfnamefont{P.}~\bibnamefont{Schiff}}, \bibnamefont{and}
  \bibinfo{author}{\bibfnamefont{A.}~\bibnamefont{Nitzan}},
  \bibinfo{journal}{Phys. Chem. Chem. Phys.} \textbf{\bibinfo{volume}{14}},
  \bibinfo{pages}{13746} (\bibinfo{year}{2012}),
  \urlprefix\url{http://dx.doi.org/10.1039/C2CP41442B}.

\bibitem[{\citenamefont{Kirkwood}(2004)}]{kirkwood_statistical_2004}
\bibinfo{author}{\bibfnamefont{J.~G.} \bibnamefont{Kirkwood}},
  \bibinfo{journal}{J. Chem. Phys.} \textbf{\bibinfo{volume}{3}},
  \bibinfo{pages}{300} (\bibinfo{year}{2004}).

\bibitem[{\citenamefont{Talkner and Hänggi}(2020)}]{talkner_colloquium_2020}
\bibinfo{author}{\bibfnamefont{P.}~\bibnamefont{Talkner}} \bibnamefont{and}
  \bibinfo{author}{\bibfnamefont{P.}~\bibnamefont{Hänggi}},
  \bibinfo{journal}{Rev. Mod. Phys.} \textbf{\bibinfo{volume}{92}},
  \bibinfo{pages}{041002} (\bibinfo{year}{2020}).

\bibitem[{\citenamefont{Bruch et~al.}(2016)\citenamefont{Bruch, Thomas,
  Viola~Kusminskiy, von Oppen, and Nitzan}}]{bruch_quantum_2016}
\bibinfo{author}{\bibfnamefont{A.}~\bibnamefont{Bruch}},
  \bibinfo{author}{\bibfnamefont{M.}~\bibnamefont{Thomas}},
  \bibinfo{author}{\bibfnamefont{S.}~\bibnamefont{Viola~Kusminskiy}},
  \bibinfo{author}{\bibfnamefont{F.}~\bibnamefont{von Oppen}},
  \bibnamefont{and} \bibinfo{author}{\bibfnamefont{A.}~\bibnamefont{Nitzan}},
  \bibinfo{journal}{Phys. Rev. B} \textbf{\bibinfo{volume}{93}},
  \bibinfo{pages}{115318} (\bibinfo{year}{2016}).

\bibitem[{\citenamefont{Ochoa et~al.}(2016)\citenamefont{Ochoa, Bruch, and
  Nitzan}}]{ochoa_energy_2016}
\bibinfo{author}{\bibfnamefont{M.~A.} \bibnamefont{Ochoa}},
  \bibinfo{author}{\bibfnamefont{A.}~\bibnamefont{Bruch}}, \bibnamefont{and}
  \bibinfo{author}{\bibfnamefont{A.}~\bibnamefont{Nitzan}},
  \bibinfo{journal}{Phys. Rev. B} \textbf{\bibinfo{volume}{94}},
  \bibinfo{pages}{035420} (\bibinfo{year}{2016}).

\end{thebibliography}

\end{document}